\documentclass[11pt]{article}
\usepackage{graphicx,epsfig}

\newcommand{\BABARPubYear}    {00}

\newcommand{\BABARProcNumber} {35}
\newcommand{\SLACPubNumber} {8724}

\newcommand {\vtd}      {V_{td}}
\newcommand {\vts}      {V_{ts}}
\newcommand {\dstlnu} {D^{\ast} \ell \bar \nu}
\newcommand {\dstpi} {D^{\ast} \pi}
\newcommand {\dststlnu} {B^{\pm} \rightarrow D^{\ast} X \ell \bar \nu}
\newcommand {\wtag} {w_{tag}}
\newcommand {\chideff} {\chi_{d}^{\mathrm{eff}}}

\def\ssb  {\sin 2 \beta }
\def\dstar {D^\ast}
\def\bi{\begin{itemize}}
\def\ei{\end{itemize}}
\def\ra{\rightarrow}

\def\ps{\mathrm {ps}}
\def\deltamd{\Delta m_{B_d}}
\def\bz{B^0}
\def\bzb{\overline{B}{}^0} %<<<

\def\lbabar{\mbox{{\large\sl B}\hspace{-0.4em} {\normalsize\sl A}\hspace{-0.03em}{\large\sl B}\hspace{-0.4em} {\normalsize\sl A\hspace{-0.02em}R}}}
\usepackage{relsize}
\def\babar{\mbox{\slshape B\kern-0.1em{\smaller A}\kern-0.1em
    B\kern-0.1em{\smaller A\kern-0.2em R}}}

\def\Kbar  {\kern 0.2em\overline{\kern -0.2em K}{}}

\def\Kzb   {\ensuremath{\Kbar^0}}
\def\KzKzb {\ensuremath{K^0 \kern -0.16em \Kzb}}

\def\Dbar  {\kern 0.2em\overline{\kern -0.2em D}{}}

\def\Dzb   {\ensuremath{\Dbar^0}}
\def\DzDzb {\ensuremath{D^0 {\kern -0.16em \Dzb}}}

\def\Bbar  {\kern 0.18em\overline{\kern -0.18em B}{}}

\def\Bzb   {\ensuremath{\Bbar^0}}

\def\BzBzb {\ensuremath{B^0 {\kern -0.16em \Bzb}}}

\mathchardef\Upsilon="7107
\def\Y#1S{\ensuremath{\Upsilon{(#1S)}}}% no space before {...}!

\mathchardef\Deltares="7101
\mathchardef\Xi="7104
\mathchardef\Lambda="7103
\mathchardef\Sigma="7106
\mathchardef\Omega="710A
\def\Deltabar   {\kern 0.25em\overline{\kern -0.25em \Deltares}{}}
\def\Lbar {\kern 0.2em\overline{\kern -0.2em\Lambda\kern 0.05em}\kern-0.05em{}}
\def\Sigbar{\kern 0.2em\overline{\kern -0.2em \Sigma}{}}
\def\Xibar{\kern 0.2em\overline{\kern -0.2em \Xi}{}}
\def\Obar{\kern 0.2em\overline{\kern -0.2em \Omega}{}}
\def\Nbar{\kern 0.2em\overline{\kern -0.2em N}{}}
\def\Xbar{\kern 0.2em\overline{\kern -0.2em X}{}}

\def\ev   {\ensuremath{\rm \,e\kern -0.08em V}}
\def\kev  {\ensuremath{\rm \,ke\kern -0.08em V}} 
\def\mev  {\ensuremath{\rm \,Me\kern -0.08em V}} 
\def\gev  {\ensuremath{\rm \,Ge\kern -0.08em V}} 
\def\gevc {\ensuremath{{\rm \,Ge\kern -0.08em V\!/}c}} 
\def\tev  {\ensuremath{\rm \,Te\kern -0.08em V}}
\def\mevc {\ensuremath{{\rm \,Me\kern -0.08em V\!/}c}} 
\def\gevcc{\ensuremath{{\rm \,Ge\kern -0.08em V\!/}c^2}} 
\def\mevcc{\ensuremath{{\rm \,Me\kern -0.08em V\!/}c^2}}

\def\fb   {\ensuremath{\mbox{\,fb}}}

\def\mus  {\ensuremath{\rm \,\mus}}

\def\ps   {\ensuremath{\rm \,ps}}

\def\mus        {\ensuremath{\,\mu{\rm s}}}    %% microsecond
\def\ps         {\ensuremath{{\rm \,ps}}}   %% picosecond
\def\gsim{{~\raise.15em\hbox{$>$}\kern-.85em
          \lower.35em\hbox{$\sim$}~}}
\def\lsim{{~\raise.15em\hbox{$<$}\kern-.85em
          \lower.35em\hbox{$\sim$}~}}

\def\ra                 {\ensuremath{\rightarrow}}

\def\pep2{PEP-II}

\def\deltamd{\ensuremath{{\rm \Delta}m_d}}

\providecommand{\eqref}[1]{Eq.~(\ref{eq:#1})}

\def\jetset74   {\mbox{\tt Jetset \hspace{-0.5em}7.\hspace{-0.2em}4}}

\def\fb{\mathrm {fb}^{-1}}
\long\def\inst#1{\par\nobreak\kern 4pt\nobreak
    {\it #1}\par\vskip 10pt plus 3pt minus 3pt}

\begin{document}
{\pagestyle{empty}

\begin{flushright}
SLAC-PUB-\SLACPubNumber \\
\babar-PROC-\BABARPubYear/\BABARProcNumber \\
%\babar-PUB-\BABARPubYear/\BABARPubNumber \\
%hep-ex/\LANLNumber \\
November, 2000 \\
\end{flushright}

\par\vskip 4cm

% Title of the paper
\begin{center}
\Large \bf Physics results from \babar\ and prospects.
\end{center}
\bigskip

\begin{center}
\large 
Marie-H\'el\`ene Schune \\
Laboratoire de l'Acc\'el\'erateur Lin\'eaire \\
Batiment 208, BP 34 \\
91898 Orsay Cedex, France \\
(for the \lbabar\ Collaboration)
\end{center}
\bigskip \bigskip

% Abstract
\begin{center}
\large \bf Abstract
\end{center}
The \babar\ experiment has been taking data since about one year. 
Physics results, complementary to 
those described in \cite{gab}, are presented :
various measurements of the neutral and charged B mesons lifetimes, 
the $\bz$ mixing frequency, the mistag fractions, and the 
$B\ra D_s^{(\ast)} X$ and $\bz \ra D_s^{(\ast)+} D^{\ast - }$ 
branching ratios. 
Some prospects for the close future are shown. All the numbers 
given here are preliminary. 
\vfill
\begin{center}
Contribued to the Proceedings of the International Conference On CP Violation Physics, \\
9/18/2000---9/22/2000, Ferrara, Italy
\end{center}

\vspace{1.0cm}
\begin{center}
{\em Stanford Linear Accelerator Center, Stanford University, 
Stanford, CA 94309} \\ \vspace{0.1cm}\hrule\vspace{0.1cm}
Work supported in part by Department of Energy contract DE-AC03-76SF00515.
\end{center}

\section{Physics results}
The results shown here have been chosen not only
for their interest but also because they are representative of future 
measurements : 

\bi
\item the B lifetime measurement using inclusive reconstruction is a prototype analysis 
for CP violation studies using the same type of technique. 
\item the B lifetime measurement using fully reconstructed B decays
demonstrates the good vertex resolution of the Silicon Vertex Tracker. 
\item the $\deltamd$ measurement using dileptons events relies heavily 
both on the  vertex resolution  and the lepton identification
\item when this measurement is performed using fully reconstructed 
B decays it can be really seen as a practice ground for CP violation
 analyses
\item finally the measurements of BR($B\ra D_s^{(\ast)} X$) and 
BR$(\bz \ra D_s^{(\ast)+} D^{\ast - })$ are steps towards a 
better understanding of the B decays. 
\ei  
\subsection{ B lifetimes}
The $\bz$ and charged B lifetimes can be measured in \babar\ due to the boost 
of the $\Upsilon(4S)$ : the two B mesons are 
separated by an average distance in z of $ \sim 260 \ \mu m$. 
\par
The first result given here is 
performed using an inclusive reconstruction of the $\bz$.
An integrated luminosity of $7.9 \ \fb$ has been used. 
The $\bz$ decay 
 which is considered is $\bzb \ra D^{*+} \pi_f^-$ where 
the $D^{*+}$ is signed only by 
the soft pion coming from the decay $D^{*+} \ra D^0 \pi_s^+$. This 
soft pion is combined with the fast bachelor one ($\pi_f^-$). There are
enough constraints  to reconstruct the $D^0$ missing
 mass, assuming that the slow and fast pions are coming from a $\bz$ decay 
into $\dstpi$~\cite{inclusive_b}. The missing mass is shown 
in Figure~\ref{fig:taub_partial_md0}. 
\begin{figure}[htbp] 
\begin{center}
  \mbox{\epsfig{file=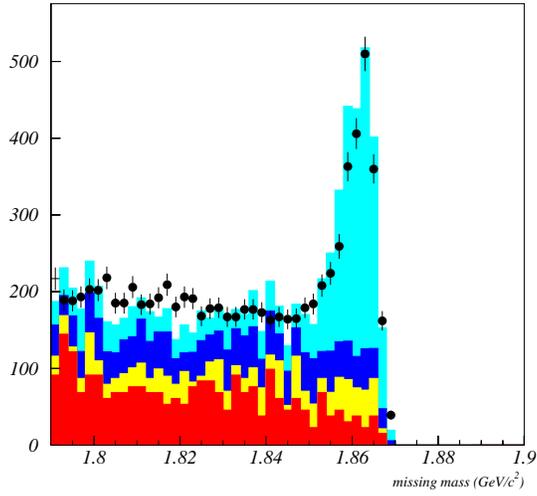,width=7.0cm}}
\end{center}
\caption{Missing mass for partially reconstructed 
$\bzb \ra D^{*+} \pi_f^-$ events. The points are the data, the shaded histograms the
Monte Carlo contributions.} 
\label{fig:taub_partial_md0}
\end{figure}

The $\bz$ lifetime is measured to be :
$$
\tau(\bz) = 1.55 \pm 0.05 \pm 0.07  \ps
$$
The main systematics are 
due to the backgrounds and the $\Delta z$ resolution function.

\par
The $\bz$ and $B^{\pm}$ lifetimes have been also measured 
using  less abundant but fully exclusive B decay modes \cite{taub_exclusif}. 
The B mesons are reconstructed using $D^{(*+)} \pi$ $D^{(*+)} \rho$, 
$D^{(*+)} a_1$ and $J \psi K^*$ decay modes. 
There is only one background source : the combinatorial background which 
is estimated from the side-bands of the beam energy substituted mass variable.  
The two proper time fits are shown in 
Figure~\ref{fig:taub_exclusif} and the results  for 
an integrated luminosity of $7.4 \fb$ are given in Table~\ref{tab:taub}. 

\begin{figure}[htbp] 
\begin{center}
  \mbox{\epsfig{file=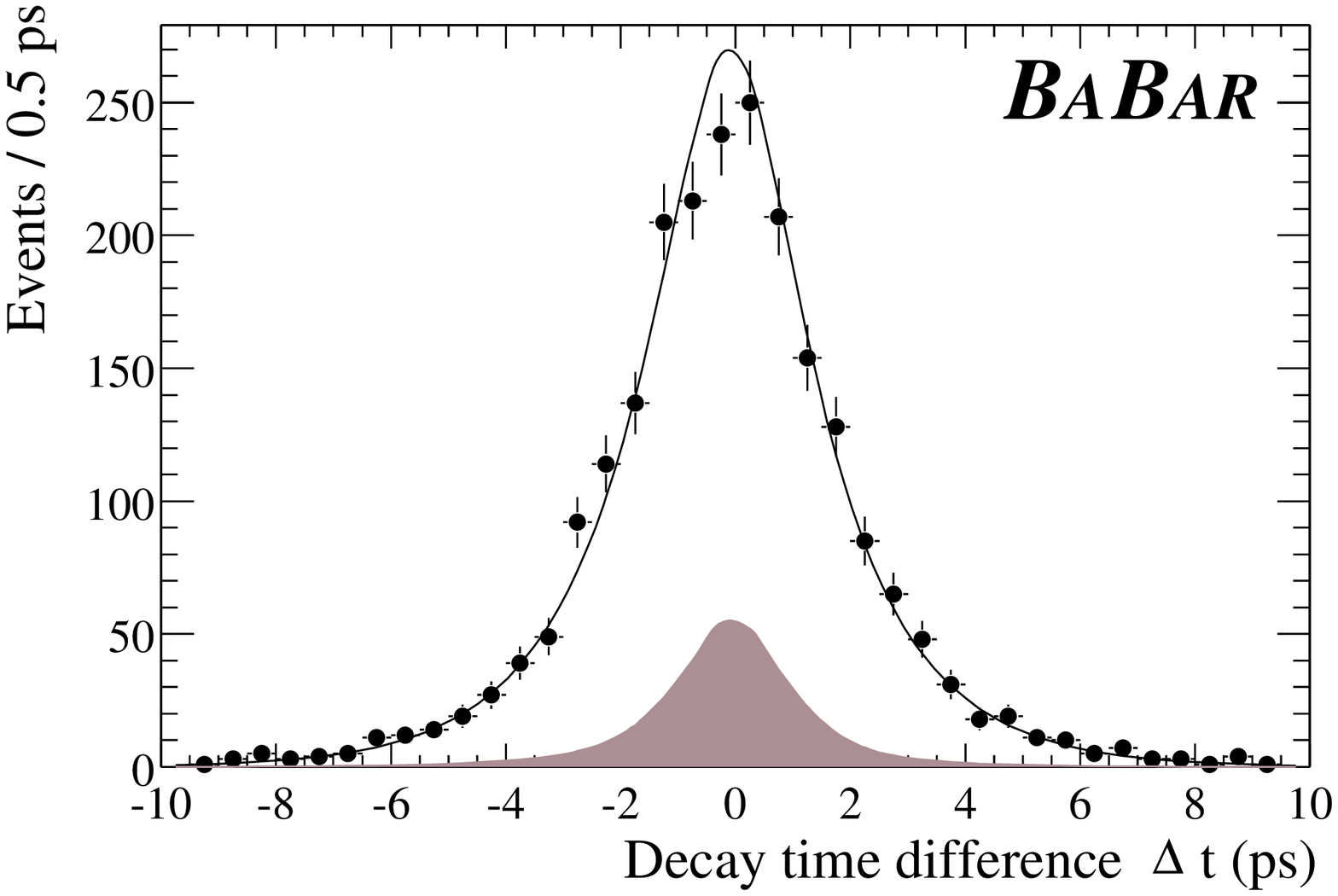,width=7.0cm}}
  \mbox{\epsfig{file=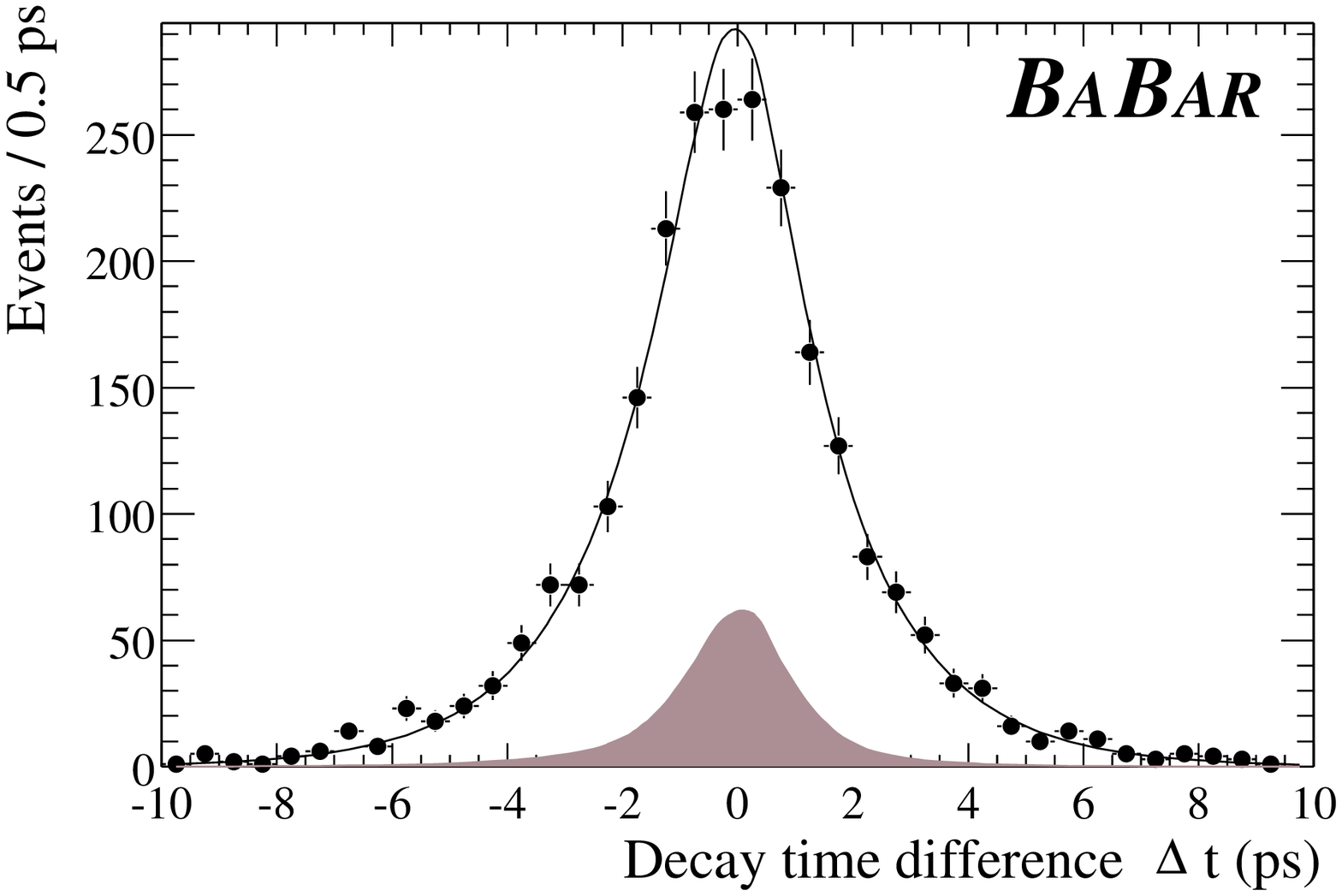,width=7.0cm}}
\end{center}
%\vskip -1.cm 
\caption{$\Delta z$ distributions for the $\bz$ (right) and charged B (left) candidates. 
The result of the lifetime fit is superimposed. The background is shown by the hatched distributions.}
\label{fig:taub_exclusif}
\end{figure}

\begin{table}[htb]
\caption{Measurements of the charged and neutral B mesons lifetimes using fully
reconstructed decay modes. The dominant systematics are related to 
the MC statistics and the background modelling.}
\label{tab:taub}
\begin{center}
\begin{tabular}{|l|l|}
\hline
$\tau(\bz)$ 		  & $1.506 \pm 0.052 \pm 0.029 \ps$ \\
\hline
$\tau(B^{\pm})$	  & $1.602 \pm 0.049 \pm 0.035  \ps$ \\
\hline
$\tau(B^{\pm})/\tau(\bz)$ & $1.065 \pm 0.044 \pm 0.021$ \\
\hline
\end{tabular}
\end{center}
\end{table}
\par
These B lifetime measurements are in good agreement with the values from 
the PDG2000 \cite{pdg}.

\subsection{$\deltamd$ measurements} 
The $\Upsilon(4S)$ resonance decays coherently into a $\bz$ $\bzb$ pair. One of these 
two B mesons  (let's take for example that the $\bz$) 
decays at time $t_1$, then the other one (the $\bzb$) starts to oscillate
and decays at time $t_2$.
If the second B decays as a $\bz$ the event will be named as {\em mixed}
and the time behaviour will follow a $e^{- | \Delta t|/\tau} 
\left(  1 -  \cos{ \deltamd  \Delta t} \right)$ law ($\Delta t = t_1 -t_2$).
If it decays as a 
$\bzb$ the event is called {\em unmixed} and the time behaviour 
will follow a $e^{- | \Delta t | /\tau} 
\left(  1 + \cos{ \deltamd  \Delta t} \right)$ law. 
The main ingredients for a $\deltamd$ measurement are :
\bi
\item to identify the flavor of one B (the one which decays 
at $t_1$)
\item to tag the flavour of the other B  (the one which decays at $t_2$)
\item to measure the distance between the two vertices in order to 
deduce $\Delta t$
\ei
In the first analysis presented here the two B mesons are tagged by two energetic leptons. 
The performances of the 
lepton identification are given in Table~\ref{table:lep_id}.
\begin{table}[htb]
\caption{Efficiencies and misidentification for the lepton used in the dilepton analysis.}
\label{table:lep_id}
\begin{center}
\begin{tabular}{|l|l|l|}
\hline
Lepton 	& Efficiency	& Mis identification \\
\hline
electron & $\sim 88 \%$ & $\sim .3 \% $ \\
muon & $\sim 75 \%$ & $\sim 3 \%$  \\
\hline
\end{tabular}
\end{center}
\end{table}
\par
The background coming from secondary leptons ($b \ra c\ra \ell$) is reduced
 by the use of a Neural Network based on 5 discriminating variables  
 (the two leptons momentum in the  $\Upsilon(4S)$ rest frame, the total energy 
in the event, the missing momentum and the angle between the 2 leptons).
The $\deltamd$ extraction is performed 
by a binned maximum likelihood fit on the asymmetry 
between like and unlike sign events. 
The fit is simultaneously done for
$\deltamd$, the sample composition (the B$^{\pm}$ contribution) and the 
mistag fraction (and its time dependence). The data and 
the fit are shown in Figure~\ref{fig:ll_res}. With an integrated 
luminosity of $7.4 \fb$, $\deltamd$ is measured to be equal to \cite{dmd_ll} :
$$
\deltamd  = (0.507 \pm 0.015 \pm 0.022) \hbar \ps ^{-1}  
$$
\begin{figure}[htbp] 
\begin{center}
  \mbox{\epsfig{file=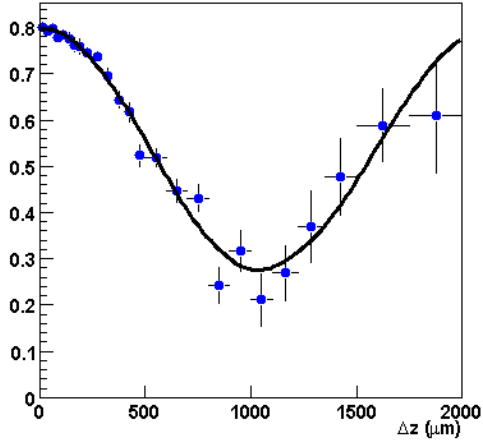,width=7.0cm}}
\end{center}
\caption{Distribution of the measured asymmetry between unlike sign ({\em unmixed}) and like sign
({\em mixed}) events. Ignoring resolution 
effects it should follow a $\cos (\deltamd \Delta t)$ law. 
The curve represents the result of the fit.}
\label{fig:ll_res}
\end{figure}
\par
The $\deltamd$ measurement using fully reconstructed $\bz$ is in fact 
divided into two samples : the hadronic sample with a $\bz$ reconstructed 
using  D$^{(\ast)} \pi$, D$^{(\ast)} \rho$, D$^{(\ast)} a_1$
and $J \psi {\mathrm K}^{\ast}$. With an integrated luminosity of $8.9 \fb$
about 2600  candidates with a purity of about $ 86 \%$ are reconstructed. 
The background is of combinatorial type and is estimated  from the 
side bands of the beam
 energy substituted mass variable .  The other sample is $\bz \ra \dstlnu$  which allows 
to select about 7500 candidates with a purity of roughly $70 \%$. The backgrounds are of different types :
	\bi
	\item combinatorial background from the $\dstar$, it is estimated from 
the $\Delta m$=M(D$^*$)-M(D$^0$) side bands
	\item fake $\ell$ ( it is obtained from control samples on the data) 
	\item $\dstar$ and $\ell$ from 2 Bs (taken mainly from data)
	\item continuum background (estimated from Off resonance data)
	\item $\dststlnu$ (estimated from LEP measurements, its 
relative efficiency is obtained from \babar\ MC)
	\ei
\par
The flavor of the other B is tagged using a prioritized algorithm :
\begin{enumerate}
\item $\ell$  : $p^{\ast}_{\ell} > 1.1 \mathrm{GeV/c}$, tag with $Q_{\ell}$ (Priority($e)>$Priority($\mu$))
\item K  :  tag with $ \sum Q_K$
\item a Neural Network for the rest of the event. The output is divided into 
two regions {\tt NT1} and {\tt NT2}.
\end{enumerate}
An unbinned maximum likelihood fit is performed on the probability distribution 
functions for the {\em mixed} and {\em unmixed} events. 
It treats 
simultaneously $\deltamd$, the main parameters of the resolution function 
and the mistag fraction (see section\ref{sec:tagmix}).

The results are displayed
 using the asymmetry between the two types of events and are shown on 
Figure~\ref{fig:dmd_excl}. The results for the hadronic and semileptonic samples are
\cite{bad81} :
\begin{eqnarray*}
\deltamd \ \mathrm{(had.)} & = (0.516 \pm 0.031 \pm 0.018) \hbar \ps ^{-1}\\
\deltamd \ \mathrm{(semilep.)}& =(0.508 \pm 0.020 \pm 0.022) \hbar \ps ^{-1}\\
\end{eqnarray*}
The combined result is :
$$
\deltamd  = (0.512 \pm 0.017 \pm 0.022) \hbar \ps ^{-1} 
$$

The main systematics are due to the $\Delta$t resolution function, the MC statistics
and the $\dststlnu$ background for the semileptonic sample. 
\par
The various $\deltamd$ measurements are in agreement with the PDG2000 \cite{pdg}. 
\begin{figure}[htbp] 
%\begin{flushleft}
%  \mbox{\epsfig{file=fully_res.eps,width=3.9cm}}
%\end{flushleft}
%\begin{flushright}
%\vskip -4.50cm 
%  \mbox{\epsfig{file=dmd_dstlnu_res.eps,width=3.9cm}}
%\end{flushright}
\vskip -1.5cm 
\begin{center}
 \mbox{\epsfig{file=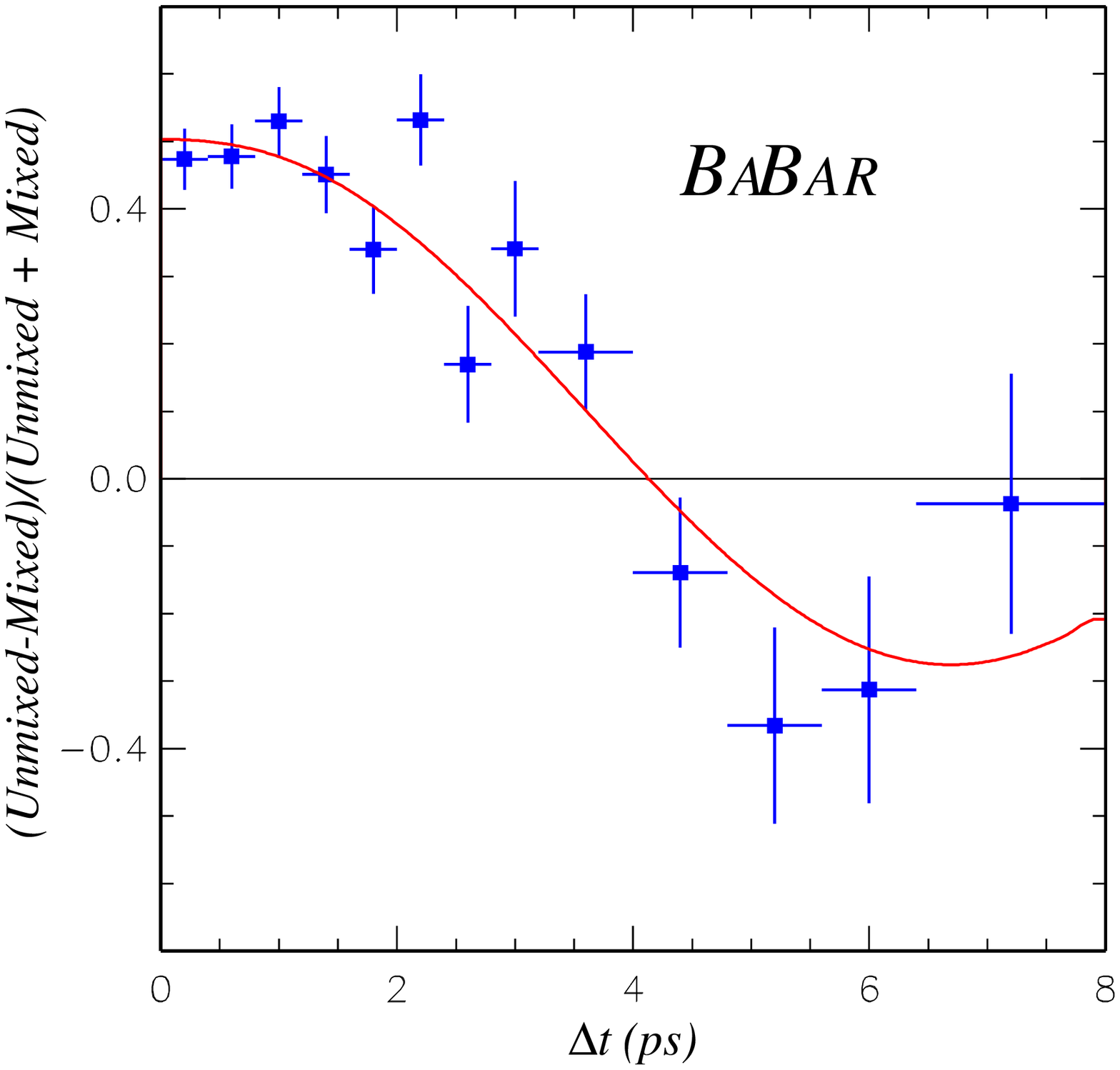,width=7.0cm}}
 \mbox{\epsfig{file=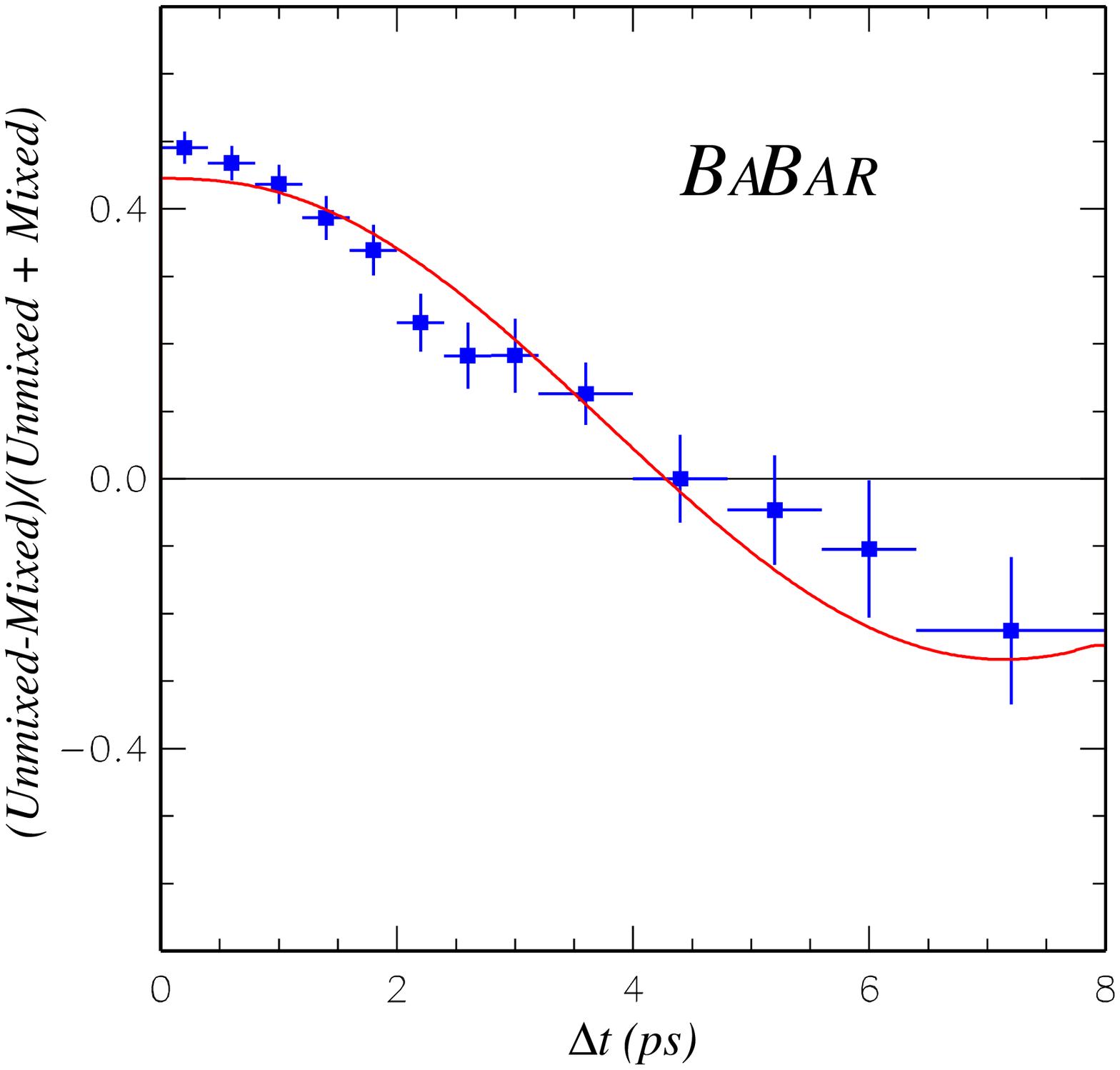,width=7.0cm}}
\end{center}
\caption{$\Delta t$ distributions in data for the hadronic (right) and semileptonic (left)
selected B decays. The fitted  $\Delta t$ shapes for the selected candidates and for the fraction of background 
are overlaid.}
\label{fig:dmd_excl}
\vskip -.5cm 
\end{figure}

\subsection{Mistag fraction measurements \label{sec:tagmix}} 
The knowledge of the mistag fractions is a pre-requisite for the analyses aiming at measuring 
$\ssb$. The mistag fraction extraction should be done on a data set 
identical to the one used for the CP analysis. It can be done simultaneously with the 
$\deltamd$ likelihood fit or on the same events but with a counting method. 
\par
For this last case, considering signal events
the fraction of {\em mixed} events can be expressed as 
$N_{\mathrm {events}}^{\mathrm {mixed}}/N_{\mathrm {events}} = \chideff  + (1 - 2 \chideff ) \wtag$
where $ \chideff$ is the mixed signal fraction within a time interval ($\Delta T$). 
This fraction is shown for charged and neutral B on Figure~\ref{fig:wtag_versust}. The oscillatory 
term is clearly visible for the neutral B. If the whole range in $\Delta T$  
is used one has $\chideff = \chi_d = 0.174 \pm 0.009$\cite{pdg}. However from  Figure~\ref{fig:wtag_versust} 
it can be seen that all the information about $\wtag$ is contained in the $\Delta t$ region
close to 0. This range can be optimized in order to reduce the statistical uncertainty 
due to the loss of information of a counting method compared to a full likelihood fit
and is shown in Figure~\ref{fig:wtag_versust}. 
The two methods are 
complementary since the counting one is simple, has a weak dependence on the $\Delta t$
resolution function (which is however obtained from data), but needs as an input 
$\deltamd$. The full likelihood fit performs simultaneously the extraction of
the mistag fractions, the main parameters of the $\Delta t$
resolution function and $\deltamd$. 
\par
The figure of merit for the tagging is 
$ Q = \varepsilon (1-2 \wtag )^2$ since $1/\sigma^2(\ssb) \propto \varepsilon (1-2 \wtag )^2$. 
The results are given in Table~\ref{tab:wtag} for 
an integrated luminosity of 8.9 $\fb$\cite{bad81}.

The overall quality factor $Q$ is of the order of 28\%. The results of the two methods 
(full likelihood and counting) are in agreement. The main systematics are due the 
knowledge of the backgrounds (for the $\dstlnu$~sample), the $\Delta$z resolution and the 
MC statistics. 

\begin{figure}[htb] 
\vskip -1.5cm 
\begin{center}
  \mbox{\epsfig{file=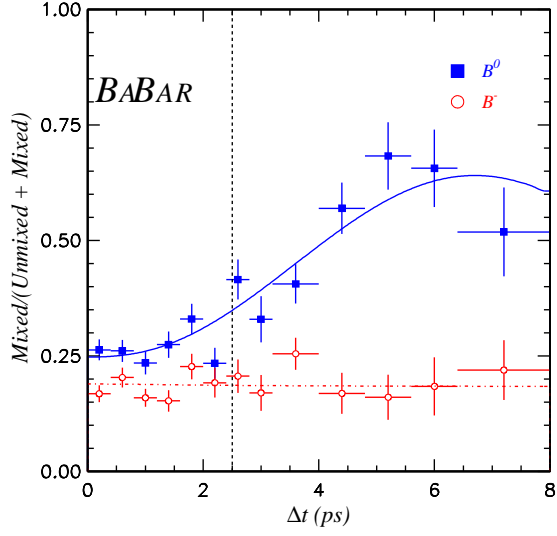,width=8.0cm}}
\end{center}
\caption{Fraction of mixed events as a function of $|\Delta t|$ for events in the 
hadronic sample,
for neutral B mesons (full squares) and charged B mesons (open circles). All tagging categories are included. 
The rate of mixed events at $\Delta t=0$ is governed by $\wtag$. The dot-dashed line at  $\Delta t$ 
=2.5 ps indicates the 
boundary of the counting method.}
\label{fig:wtag_versust} 
\end{figure}

\begin{table}[htb]
\caption{Mistag fractions and tagging efficiencies measured on the data for the four tagging categories.}
\label{tab:wtag}
\begin{center}
\begin{tabular}{|l|l|l|}
\hline
tag 		&  Mistag fraction($\wtag$) 		& $\varepsilon$\\
\hline
$\ell$		&  $0.096 \pm  0.017 \pm 0.013 $  	& $0.112 \pm 0.005$\\
K		&  $0.197 \pm  0.013 \pm 0.011$   	& $0.367 \pm 0.009$ \\
{\tt NT1}   	&  $0.167 \pm  0.022 \pm 0.020$   	& $0.117 \pm 0.005$ \\
{\tt NT2}   	&  $0.331 \pm  0.021 \pm 0.021$   	& $0.166 \pm 0.006$\\
\hline
\end{tabular}
\end{center}
\end{table}

\subsection{BR($B\ra D_s^{(\ast)} X$) and BR$(\bz \ra D_s^{(\ast)+} D^{\ast - })$} 
With an integrated luminosity of 
$7.7 \fb$ about 19000  $D_s$ are reconstructed in the $\phi \pi (\phi \ra \pi^+ \pi^-)$ mode. 
The invariant mass plot is shown in Figure~\ref{fig:ds_serguey} where both the $D_s$ 
and the Cabibbo suppressed decay $D \ra \phi \pi$ are clearly visible. The branching ratio
BR($ B\ra D_s X)$ is extracted\cite{serguey1}:
$$(11.9 \pm 0.3 \pm 1.1 \pm 3.0)\%$$ 
The last and dominating uncertainty is due to the bad knowledge of the branching ratio of the 
decay $D_s \ra \phi \pi$\cite{pdg}. Once a $D_s$ is found it can be paired with photons in order 
to try to reconstruct the decay $D_s^* \ra D_s \gamma$. The Figure~\ref{fig:ds_serguey} shows
the mass difference between the $D_s^*$ and the $D_s$. The branching ratio
$$ \mathrm{BR}(B\ra D_s^{\ast} X) = (6.8 \pm 0.7 \pm 0.8 \pm 1.7)\%$$
is obtained\cite{serguey1}. 
Both measurements are in agreement with the PDG2000 \cite{pdg}. 
The branching ratios for the decay modes $\bz \ra D_s^{(\ast)+} D^{\ast - }$
are measured using a partial reconstruction technique very similar to that used for the 
$\tau(\bz)$ analysis : the $D^*$  is signed by the soft pion coming from its decay. 
The results are\cite{serguey1} :
$$
{\mathrm BR}(\bz \ra  D_s^+ D^{\ast -}) = (7.1 \pm 2.4 \pm 2.5 \pm 1.8) \ 10^{-3}\\
$$
$$
{\mathrm BR}(\bz \ra  D_s^{\ast+} D^{\ast -}) = (2.5 \pm 0.4 \pm 0.5 \pm 0.6) \ 10^{-2}\\
$$
The last systematical uncertainties are due to the mis-knowledge of the 
$D_s \ra \phi \pi$ branching ratio. 
\begin{figure}[htbp] 
\begin{flushleft}
  \mbox{\epsfig{file=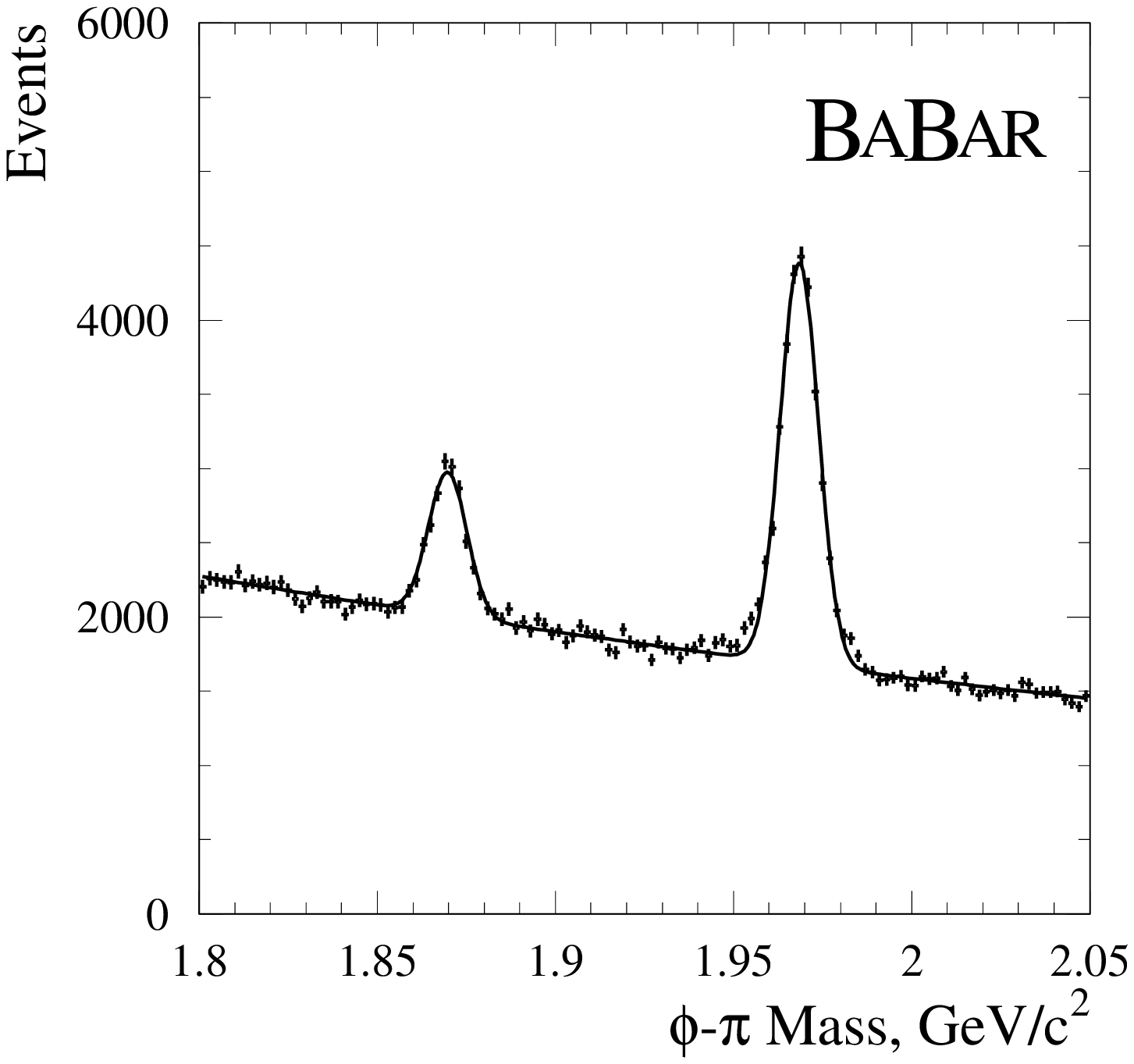,width=7.1cm}}
\end{flushleft}
\begin{flushright}
\vskip -7.45cm 
  \mbox{\epsfig{file=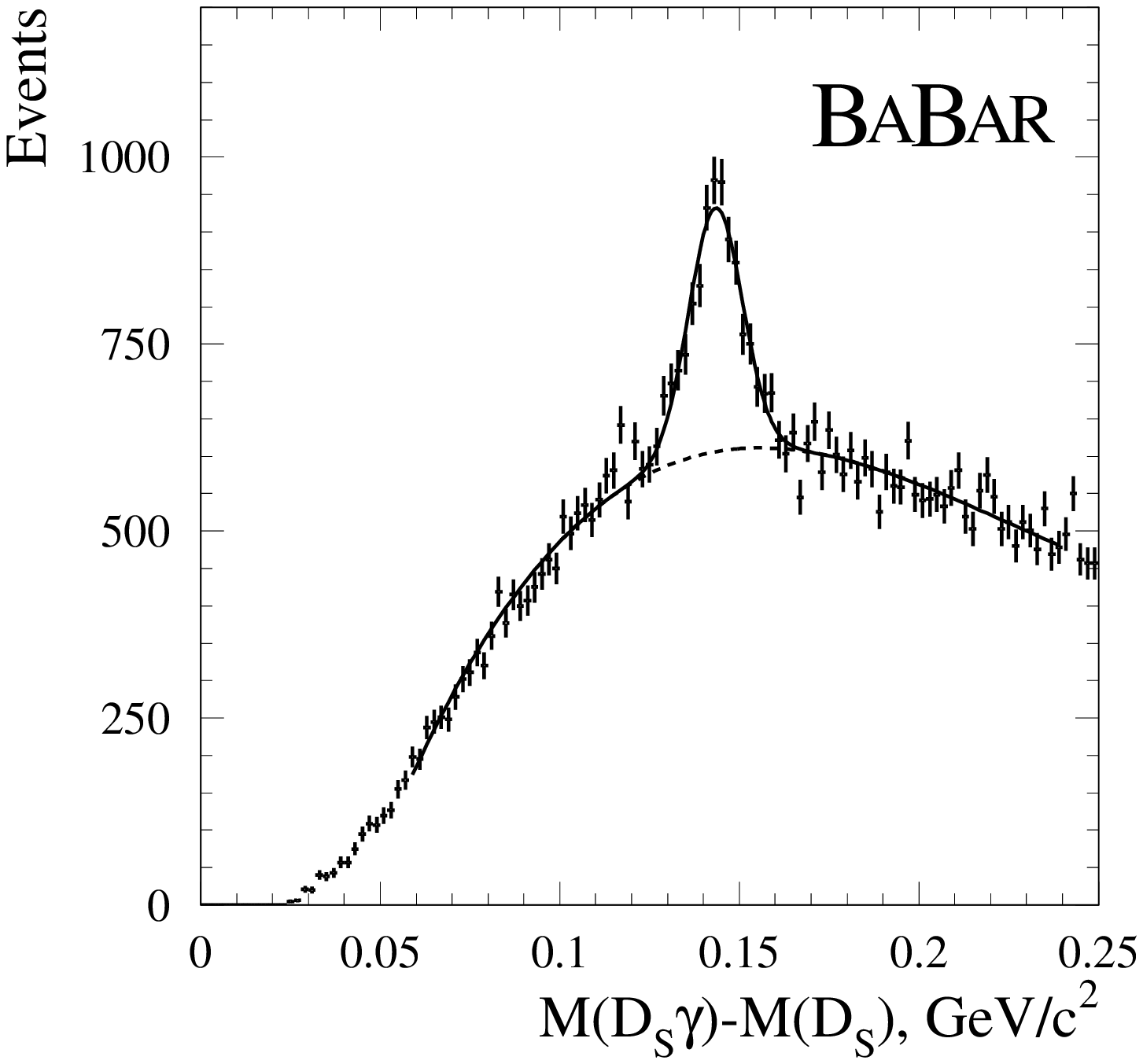,width=7.1cm}}
\end{flushright}
\vskip -.5cm 
\caption{Left : $\phi \pi$ invariant mass spectrum, both the $D_s$ and the $D$ 
(Cabibbo suppressed) peaks are clearly visible. Right : $M_{D_s^* \gamma}-M_{D_s^*}$ mass spectrum. 
The fit function is a single Gaussian for the signal and a third-order polynomial for the background.}
\label{fig:ds_serguey}
\end{figure}

\section{Prospects}
In view of drawing some prospects it is useful to note that 
the {\tt PEP-II} machine is very well working : since June 1999 an integrated 
luminosity of 19.5 $\fb$ has delivered out of which  18 $\fb$  have been 
recorded by \babar\ . In order to be able to make some previsions 
some anticipated evolutions have been done and are given in 
Table~\ref{tab:pep2}~\cite{detector_upgrade}. 
\begin{table}[htbp]
\caption{Anticipated performances of the {\tt PEP-II} accelerator for the coming years.}
\label{tab:pep2}
\begin{center}
\begin{tabular}{|l|l|l|}
\hline
Year &         Peak Lumi     &   $\int L dt \  (\fb)$ \\
     &($10^{33} cm^2s^{-1}$) &   total \\
\hline
 2000 &                2     & 25 \\
 2001  &               5     &  65 \\
 2002   &              8     &  145 \\
 2003   &             10     &  260 \\
\hline
\end{tabular}
\end{center}
\end{table}
For the purpose of the prospects  I will stop at the end of the year 2001 
which corresponds to an integrated luminosity of 65 $\fb$. No major 
detector upgrades are expected to happen before 2002. In 2002 trigger 
improvements will be necessary (z cuts at the level 1 trigger) in order to 
cope with the increased currents in the two beams (and backgrounds!). In 
addition the modules of the Silicon Vertex Tracker of the horizontal plane 
will be replaced because of the higher radiation level in that region. Finally 
the DIRC electronics will be modified. 
Drawing some prospects is not an easy task \dots only some examples are 
going to be shown.
\subsection{$\ssb$ prospects}
With an integrated luminosity of $9 \ \fb$ $\ssb$ has been measured to be \cite{ssb}
$$\ssb = .12 \pm .37 \pm .09$$\footnote{From Monte Carlo studies the average statistical 
uncertainty for the statistics used is expected to be about 0.32}. The list of 
the decay modes and the reconstructed number of events are given in Table~\ref{tab:stat_ssb}. 
\begin{table}[htbp]
\caption{Decay modes and number of events used for the $\ssb$ analysis.}
\label{tab:stat_ssb}
\begin{center}
\begin{tabular}{|l|c|}
\hline
Mode 	& N(events) \\
\hline
$\bz \ra J \psi K^0_s$ ($K^0_s \ra \pi^+\pi^-$) & $124 \pm 12$ \\
$\bz \ra J \psi K^0_s$ ($K^0_s \ra \pi^0\pi^0$) & $18 \pm 4$ \\
$\bz \ra \psi(2S) K^0_s$ ($K^0_s \ra \pi^+\pi^-$) & $27 \pm 6$\\
\hline
\end{tabular}
\end{center}
\end{table}
Out of these  170 events, 120 are tagged. With the expected integrated luminosity one should
reach a statistical uncertainty of about 0.2 at the end of 2001 ($25 \ \fb$) and about 0.12 
at the end of 2002 ($65 \ \fb$). In addition, an important part of the systematical uncertainty
is due to the statistics of the control samples and as such should decrease in the 
future.  
In addition other channels are going to be used to measure $\ssb$ : 
\bi 
\item  $\bz \ra J\psi K^0_L$ : a signal is already clearly visible in the data 
(Figure~\ref{fig:mb0_psikl}) and the yield is in agreement with the Monte Carlo 
expectation\cite{jpsi_signal}.
\begin{figure}[htb] 
\begin{center}
  \mbox{\epsfig{file=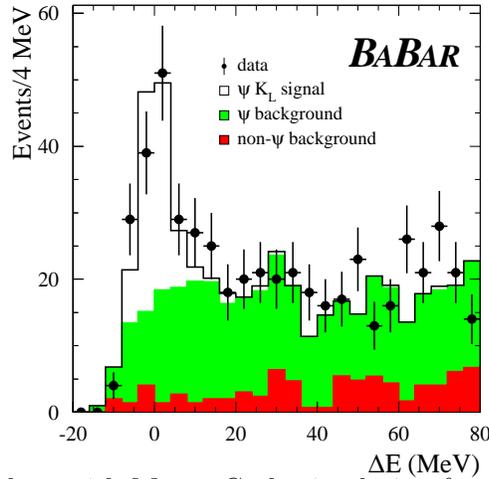,width=7.0cm}}
\end{center}
\vskip -1.5cm 
\caption{Comparison of data with Monte Carlo simulation for the $\bz \ra J\psi K^0_L$
decay.}
\label{fig:mb0_psikl}
\end{figure}
\item  $\bz \ra J\psi K^{\ast 0}$, with $K^{\ast 0} \ra K^0_s \pi^0$ : the branching ratio
is measured :$(13.8 \pm 1.1 \pm 1.8) \ 10^{-4}$ and an angular analysis using 
the charged decay mode of the $K^{\ast 0}$ confirms the fact that it is  mainly CP even so 
that the dilution of the  CP asymmetry is going to be small. 
\item $\bz \ra D^{(*)+}D^{(*)-}$ : an analysis with partial reconstruction (one of the 
$D^{*+}$ is reconstructed only via its $\pi^+_s$ and the other is fully reconstructed). This 
type of analysis has been  exercized with the $\tau(\bz)$ measurement using
$\pi_f^+  \pi^-_s$. From a full reconstruction in 
the $D^{*+}D^{*-}$ mode 30 to 40 events are expected for $25 \ \fb$. 
\ei
This list is not exhaustive and other modes are also pursued 
({\em eg} $\bz \ra \phi  K^0_s$) since it is interesting to try to measure separately $\ssb$ 
with as many modes as possible in order to search for New Physics. 
%----------------------------------------------
\subsection{Charmless B decays prospects}
As was shown at this conference by M. Neubert\cite{neubert}, with the 
measurement of a large 
number of charmless B decays, it should be possible to extract $\gamma$. 
In addition $\bz \ra \pi^+ \pi^-$ and $\bz \ra \pi^+ \pi^- \pi^0$ should allow to 
measure the angle $\alpha$. Today, with an integrated luminosity of about 8 $\fb$, the 
$\pi^+ \pi^-$, $K^+ \pi^-$ branching ratios are measured with a precision of 25 to
30 \%, and a 90 \% CL limit is obtained for $K^+K^-$\cite{charmless1}. The results are given in 
Table~\ref{tab:charmless} and compared with the results from CLEO
\cite{charmless_cleo} and BELLE\cite{charmless_belle}. 
\begin{table}[htbp]
\caption{Measurements or 90 \% CL limit for the $\bz \ra \pi^+ \pi^-, K^+ \pi^-$ and $K^+ K^-$ decay modes. 
The values are expressed in units of $ 10^{-6}$. 
The results of \babar\, CLEO and BELLE are given.}
\label{tab:charmless}
\begin{center}
\begin{tabular}{|l|l|l|l|}
\hline
Mode 	&  \babar\ & BELLE &  CLEO\\
\hline
$\pi^+ \pi^-$	& $(9.3 ^{+2.6 +1.2 }_{-2.3 -1.4 })  $ &$(6.3 ^{+3.9}_{-3.5}\pm 1.6) $ &$(4.3 ^{+1.6}_{-1.4}\pm 0.5) $  \\
$K^+ \pi^-$	& $(12.5 ^{+3.0 +1.3 }_{-2.6 -1.7} )$  &$(17.4 ^{+5.1}_{-4.6}\pm 3.4)$ &$(17.2 ^{+2.5}_{-2.4}\pm 1.2) $\\
$K^+ K^-$	& $< 6.6  $ at 90\% CL  &$< 6.0  $ at 90\% CL & $< 1.9  $ at 90\% CL \\
\hline
\end{tabular}
\end{center}
\end{table}

At the end of 2001 the statistical 
uncertainty should be reduced by a factor 2.5 and the region of theoretical interest 
should be reached. Concerning $\bz \ra \pi^+ \pi^- \pi^0$ up to now 
only a branching ratio has been measured \cite{charmless2} : 
$$\mathrm{BR}(\bz \ra \rho \pi) = (48.5 \pm 13.4 ^{+5.8}_{-5.2}) \ 10^{-6}$$. 
There is not at the present stage enough experimental information to be able to draw some 
precise prospects on the time dependent measurement of $\alpha$. 
%----------------------------------------------
\subsection{$\bz \ra K^{\ast 0} \gamma$ prospects}
This decay mode is due to the penguin diagram shown in Figure~\ref{fig:b_to_sg}. 
In the Standard Model the branching ratio is expected to be of the order of 
$(3.3- 6.3) \ 10^{-5}$. It is however sensitive to new particles  in the loop and could be 
enhanced by New Physics. With an integrated luminosity of about 8 $\fb$ the branching ratio has been 
measured \cite{b_to_s_g}:  
$$
\mathrm{BR}(\bz \ra K^{\ast 0} \gamma) = (5.42 \pm .82 \pm .47) \ 10^{-5}
$$ 
The beam energy substituted mass is shown in Figure~\ref{fig:b_to_sg}. 
At the end of 2001, with 65 $\fb$, this decay mode should be precisely measured. However, the
measurement will probably be systematically limited at that time and one will have
to perform more inclusive analyses ($b \ra s \gamma$) or search for higher resonance.  
The decay mode $B \ra \rho \gamma$ will also be searched for in order to compare it 
to $ K^{\ast} \gamma$ to obtain some information on  $\vtd/ \vts$. 
\begin{figure}[htbp] 
\begin{flushleft}
  \mbox{\epsfig{file=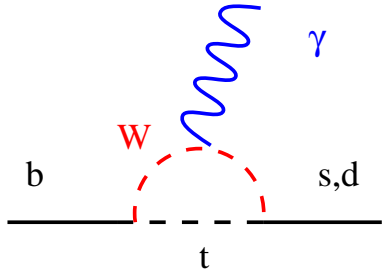,width=5.1cm}}
\end{flushleft}
\begin{flushright}
\vskip -4.cm 
  \mbox{\epsfig{file=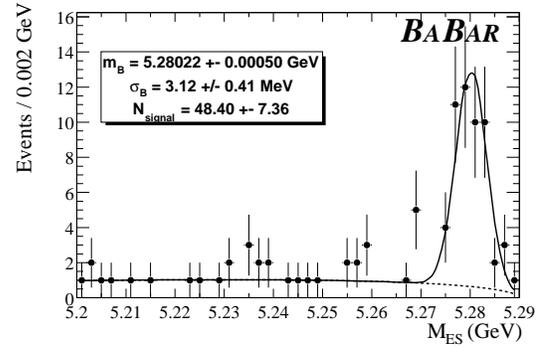,width=7.1cm}}
\end{flushright}
\vskip -.5cm 
\caption{Left : penguin diagram at the origin of the $\bz \ra K^{\ast 0} \gamma$ decay. 
Right : beam energy substituted mass projection for the $\bz \ra K^{\ast 0} \gamma$, $K^{\ast 0} \ra K^+ \pi^-$
decay.}
\label{fig:b_to_sg}
\end{figure}

\section{Conclusion}
The measurements of the Unitarity Triangle are the first priority for \babar\ not 
only to validate the Standard Model but also to search  for New Physics. This will be done 
with the measurements of the same Unitarity Triangle angle with as many modes as possible. These 
measurements have already started for the easiest modes, and the first building 
blocks are in place in the other cases. 
\par
Not only the ``sparkling'' analyses are important (and done!) : some competitive physics 
results  have already been obtained by the \babar\ collaboration. In the near future we will 
have in hand a clean sample of B mesons of an unprecedented size and we will 
use them in order to take part in the understanding of the overall picture of the B decays. 
\par
Due to lack of time I have not presented any prospects on charm and $\tau$ physics but these
two important subjects are studied also by the   \babar\ collaboration.
\par
Finally I would like to add that recording, reconstructing and analysing an integrated 
luminosity of 65 $\fb$ is both a hardware and a software challenge. 
\vskip 1cm 
\subsection*{Acknowledgements}
I would like to thank all my \babar~colleagues who helped me in preparing this
talk, in particular G. Bonneaud for the informations on the upgrade. 
Finally many thanks to St\'ephane Plaszczynski for the careful reading 
of this proceedings.


\begin{thebibliography}{99}
\bibitem{gab} G. Sciolla, First results from \babar, 2000 CP Physics Conference, Ferrara, September 2000. 
\bibitem{inclusive_b} \babar\ Collaboration,SLAC-PUB-8531. 
\bibitem{taub_exclusif} \babar\ Collaboration,SLAC-PUB-8529.
\bibitem{pdg} D.E. Groom et al, The European Physical Journal C15, 1 (2000).
\bibitem{dmd_ll} \babar\ Collaboration,SLAC-PUB-8532.
\bibitem{bad81} \babar\ Collaboration,SLAC-PUB-8530.
\bibitem{serguey1} \babar\ Collaboration,SLAC-PUB-8535.
\bibitem{detector_upgrade} \babar\ Collaboration, Detector Upgrades, 
http://www.slac.stanford.edu/BFROOT/www/ \\
Detector/Upgrades/FINALREPORT.pdf
\bibitem{ssb} \babar\ Collaboration,SLAC-PUB-8540.
\bibitem{jpsi_signal} \babar\ Collaboration,SLAC-PUB-8527.
\bibitem{neubert} M. Neubert, QCD factorization and CP asymmetries in Hadronic B decays, 2000 CP Physics Conference, Ferrara, September 2000. 
\bibitem{charmless1} \babar\ Collaboration,SLAC-PUB-8536.
\bibitem{charmless_cleo} R. A. Stroynowski, CLEO results on Charmless B meson decays, ICHEP2000, Osaka, July 2000.
\bibitem{charmless_belle} P, Chang, Studies of charmless hadronic decays of B mesons with Belle, ICHEP2000, Osaka, July 2000.
\bibitem{charmless2} \babar\ Collaboration,SLAC-PUB-8537.
\bibitem{b_to_s_g}\babar\ Collaboration,SLAC-PUB-8534.

\end{thebibliography}
\end{document}